\documentclass[12pt]{article}
\usepackage{latexsym}
\usepackage[cp1250]{inputenc}

\title{Global Symmetries of Noncommutative Space - time}
\author{C.Gonera$^*$,P.Kosi\'nski$^*$, P.Ma\'slanka\thanks{supported by the grant 1 P03B 02 128 of the Polish Ministry of Science.} \\ 
Department of Theoretical Physics II, Institute of Physics,
University of {\L}\'od\'z \\
Pomorska 149/153, 90 - 236 {\L}\'od\'z Poland.\\
S.Giller$^*$\\
Institute of Physics, Jan D{\l}ugosz Academy,\\
Armii Krajowej 13/15, 42-200 Cz\c{e}stochowa, Poland}
\date{}

\begin{document}
\maketitle
\begin{abstract}
The global counterpart of infinitesimal symmetries of noncommutative space-time is discussed.
\end{abstract}

\newpage
Some atttention has been paid recently to the problem of space-time symmetries for noncommutative field theories (NCFT), 
\cite{b1}$\div $\cite{b9}. It is fairly obvious that the full Poincare symmetry, if understood in the traditional way, 
cannot survive the "quantization" procedure consisting in replacing the commuting coordinates by the ones obeying
\begin{eqnarray}
[x^{\mu },x^{\nu }]=i\Theta ^{\mu \nu } \label{w1}
\end{eqnarray}

On the contrary, only the semidirect product of the stability subgroup of $\Theta ^{\mu \nu }$\ and the translations remains. 
However, when discussing the properties of NCFT one has often refer to the whole Poincare group. It has been suggested 
\cite{b1}, \cite{b2}, that the noncommutative space-time carries as rich symmetry as the Poincare one provided this symmetry 
is understood in the quantum group sense. On the infinitesimal level the following picture emerges,\cite{b1} - \cite{b9}.
 The 
generators of the symmetry transformations obey standard commutation rules defining the appropriate (Poincare, Weyl, 
conformal, etc.) Lie algebra. The coproduct is, however, modified by demanding that the action of Lie algebra generators 
commutes with the star product operation. Since the latter is defined with help of $\Theta ^{\mu \nu }$\ tensor, the 
coproduct acquires a nonstandard form. In order to get the explicit form of it one transforms first the star product into 
the standard one and acts with the symmetry generators (which means the standard coproduct implied by Leibniz rule) and 
finally expresses the result back in the form of star product. As a result one arrives at the deformed structure obtained 
by applying the twist $exp(\frac{i}{2}\Theta ^{\mu \nu }P_{\mu }\otimes  P_{\nu })$. One can show that resulting nice algebraic 
structure is rich enought to explain some properties of NCFT which cannot be understood of in terms of stability subgroup of 
$\Theta ^{\mu \nu }$, \cite{b1}, \cite{b2}. 

Having defined the deformation of Lie algebra one can look for the dual Hopf algebra, i.e. the deformation of the corresponding 
group. This is quite simple and can be done for example, within FRT formalism \cite{b10}. The resulting structure shares the 
coproduct with the corresponding classical group while the algebraic sector appears to be deformed. For the Poincare group 
this has been done in Ref. \cite{b8} leading to the quantum group alredy known \cite{b11}, \cite{b12}. 

One can ask what is the meaning of the nontrivial algebraic structure of symmetry group. In Ref. \cite{b9} we have shown,
 again 
in the case of Poincare group, that the algebraic relations between group elements arise if we demand that the group action 
commutes with the star product also on the global level. To this end we considered the realization of deformed group in terms 
of algebra of functions on its classical counterpart endowed with the appropriate star product. It can be then shown that 
the group action commutes with star product in the sense that the following equality holds
\begin{eqnarray}
(F_1* F_2)(gx)=F_1(gx)* F_2(gx) \label{w2}
\end{eqnarray}
where on the left hand side the standard star product on noncommutative space-time is taken prior to the group action while 
on the right hand side the star product, both in group and space-time variables is taken at the end. In this note we would 
like to point out that this scheme, checked in the case of Poincare group, works in more general framework. The reason 
for doing that is that in some circumstances more general symmetries than the Poincare one become important. This includes Weyl and 
conformal groups relevant, in the standard framework, for theories with dimensionless couplings only (in particular - with 
vanishing masses) as well as the affine group which, being dynamically broken and combined with conformal one produces 
general relativity theory with gravitons being the Goldstone modes of broken symmetry \cite{b13}. 

Assume the group $G$\ contains translations as abelian subgroup; assume further, that any element of $G$\ can be written as a 
product of translation and an element of some subgroup $S\subset G$.
$$g=e^{ia^{\mu }P_{\mu }}e^{i\zeta ^aK_a}$$
where $g\in G$, $P_{\mu }$\ are translation generators while $K_a$\ - the generators of $S$. Then writting
\begin{eqnarray}
&& g''=g'g=e^{ia'^{\mu }P_{\mu }}e^{i\zeta '^aK_a}e^{ia^{\mu }P_{\mu }}e^{i\zeta ^aK_a}= \\ \nonumber
&& =e^{ia'^{\mu }P_{\mu }}(e^{i\zeta '^aK_a}e^{ia^{\mu }P_{\mu }}e^{-i\zeta '^aK_a})(e^{i\zeta '^aK_a}e^{i\zeta ^aK_a})\label{w3}
\end{eqnarray}
one finds the composition rules
\begin{eqnarray}
&& a''^{\mu }=h^{\mu }(\underline{\zeta }',\underline{a})+a'^{\mu };\;\;\; h^{\mu }(\underline{\zeta '},0)=0 \\ \nonumber
&& \zeta ''^a=l^a(\underline{\zeta },\underline{\zeta }',\underline{a}) \label{w4}
\end{eqnarray}
Moreover, $G/S$\ can be identified with space-time and the action of $G$\ reads
\begin{eqnarray}
x'^{\mu }=h^{\mu }(\underline{\zeta },\underline{x})+a^{\mu } \label{w5}
\end{eqnarray}
and $S$\ becomes the stability subgroup of the point $x^{\mu }=0$. 

Having the composition law (\ref{w4}) one can easily calculate the left - and right - invariant fields corresponding to the 
generators $P_{\mu }$. One obtains
\begin{eqnarray}
&& P^L_{\mu }=\frac{\partial h^{\nu }(\underline{\zeta },\underline{a})}{\partial a^{\mu }}\mid _{a=0}\frac{\partial }
{\partial a^{\nu }}+\frac{\partial l^a(\underline{\zeta },0,\underline{a})}{\partial a^{\mu }}|_{a=0}\frac{\partial }
{\partial \zeta ^a} \\ \nonumber
&& P^R_{\mu }=\frac{\partial }{\partial a^{\mu }} \label{w6}
\end{eqnarray}
Both fields are obviously related by the adjoint action of $G$. 

Let us now identify the space-time $\mathcal{M}$\ with the coset space $G/S$. The functions on $\mathcal{M}$\ are identified 
with functions 
on $G$\ constant over $S$. The standard star product on $\mathcal{M}$\ can be written as
\begin{eqnarray}
\left(F_1*F_2\right)(a)=\mu \biggl(e^{\frac{i}{2}\Theta ^{\mu \nu }P^R_{\mu }
\otimes P^R_{\nu }}F_1\otimes F_2\biggr)(a) \label{w7}
\end{eqnarray}
where $\mu $\ denotes the pointwise multiplication. 

We would like to define the star product on group manifold in such a way that eq. (\ref{w2}) holds. Let us start with the 
right-hand side of eq. (\ref{w2}). Taking the star product with respect to space-time coordinates $x^\mu $\ one has 
\begin{eqnarray}
&& F_1(gx)*F_2(gx)= \nonumber \\
&& =\sum\limits_{n=0}^{\infty }\big(\frac{i}{2}\big)^n\frac{1}{n!}\Theta ^{\varrho _1\sigma _1}...\Theta ^{\varrho _n\sigma _n}
\biggl(\partial _{\varrho _1}...\partial _{\varrho _n}F_1(gx)\biggr)
\biggl(\partial _{\sigma _1}...\partial _{\sigma _n}F_2(gx)\biggr) \label{w8}
\end{eqnarray}
where
\begin{eqnarray}
\partial _{\mu }F(gx)=\frac{\partial F(h(\underline{\zeta },\underline{x})+a)}{\partial x^{\mu }}=P^R_{\mu }(x)F(h(
\underline{\zeta };\underline{x})+a) \label{w9}
\end{eqnarray}
Here $P^R_{\mu }(x)$\ is the right-invariant vector field, as given by eq. (\ref{w6}), with translations being identifield 
with space-time coordinates $x^{\mu }$. 

Now, due to the properties of invariant fields on Lie group (see Appendix, esp. eq. (20)) one has 
\begin{eqnarray}
P^R_{\mu }(x)F(h(\underline{\zeta };\underline{x})+a)=P^L_{\mu }(\underline{\zeta };\underline{a})F(h(\underline{\zeta };
\underline{x})+a) \label{w10}
\end{eqnarray}

Therefore, the star product (\ref{w8}) can be rewritten as
\begin{eqnarray}
F_1(gx)*F_2(gx)=\mu \biggl(e^{\frac{i}{2}\Theta ^{\mu \nu }P^L_{\mu }(g)\otimes P^L_{\nu }(g)}F_1(gx)
\otimes F_2(gx)\biggr) \label{w11}
\end{eqnarray}

Consider now the left-hand side of eq. (\ref{w2}). 
Using the right-invariance of $P^R_{\mu }$\ one writes
\begin{eqnarray}
&& (F_1*F_2)(gx)= \label{w12} \\
&& =\mu \biggl(e^{\frac{i}{2}\Theta ^{\mu \nu }P^R_{\mu }(x)\otimes P^R_{\nu }(x)}F_1(x)\otimes F_2(x)\biggr)_
{\mid _{x\rightarrow gx}}=
\nonumber \\
&& =\mu \biggl(e^{\frac{i}{2}\Theta ^{\mu \nu }P^R_{\mu }(a)\otimes P^R_{\nu }(a)}F_1(gx)\otimes F_2(gx)\biggr) \nonumber
\end{eqnarray}

In order to get the relevant star product on the group manifold we compare eqs. (\ref{w11}) and (\ref{w12}) (all vector 
fields commute!) 
\begin{eqnarray}
&& (F_1*F_2)(gx)=\mu \biggl(e^{\frac{i}{2} \Theta ^{\mu \nu }(P^R_{\mu }(a)\otimes P^R_{\nu }(a)-P^L_{\mu }(g)\otimes 
P^L_{\nu }(g))}\cdot \nonumber \\
&& \cdot  e^{\frac{i}{2}\Theta ^{\mu \nu }(P^L_{\mu }(g)\otimes P^L_{\nu }(g))}F_1(gx)\otimes F_2(gx)\biggr)= 
\nonumber \\
&&= \biggl [ \sum\limits_{n=0}^{\infty }(\frac{i}{2})^n\frac{1}{n!}\Theta ^{\varrho _1\sigma _1}...\Theta ^{\varrho _n\sigma _n}
(P^R_{\varrho _1}(a)P^R_{\sigma _1}(a')-P^L_{\varrho _1}(g)P^L_{\sigma _1}(g'))... \nonumber \\
&& ...(P^R_{\varrho _n}(a)P^R_{\sigma _n}(a')-P^L_{\varrho _n}(g)P^L_{\sigma _n}(g'))\biggr]F_1(gx)* F_2(g'x)_{\mid _{
g=g'}} \label{w13}
\end{eqnarray}
where the star product on the right-hand side is taken only with respect to the x -variables. We conclude that eq. (\ref{w2}) 
holds provided the star product on group manifold is defined by the following formula
\begin{eqnarray}
(H_1 * H_2)(g)=\mu \biggl(e^{\frac{i}{2}\Theta ^{\mu \nu }(P^R_{\mu }\otimes P^R_{\nu }-P^L_{\mu }\otimes P^L_{\nu })}H_1
\otimes H_2\biggr)(g) \label{w14}
\end{eqnarray}

Having defined the star product on group manifold one can derive the deformed group algebra by expressing the pointwise 
 product in terms of the star product and taking into account the commutativity of the former. Comparying the above result 
with the infinitesimal approach \cite{b1}, \cite{b2}, \cite{b6} one concludes that the quantization is achieved by 
exponentiating the classical $r$\ -matrix. 

The above procedure, when applied to the Poincare group, gives the structure considered in Refs. \cite{b11}, \cite{b12}, 
\cite{b8}. It is straightforward to extend it to the Weyl group. The resulting structure reads.
\begin{eqnarray}
&& [\Lambda ^{\mu }_{\;\;\nu },\;\cdot\;   ]=0 \nonumber \\
&& [d,\;\cdot\;  ]=0 \nonumber \\
&& [a^{\mu },a^{\nu }]=i \Theta ^{\alpha  \beta }(\delta ^{\mu }_{\;\;\alpha }\delta ^{\nu }_{\;\;\beta }-e^{2d}\Lambda 
^{\mu }_{\;\;\alpha }\Lambda ^{\nu }_{\;\;\beta }) \nonumber \\
&& \triangle (\Lambda ^{\mu }_{\;\;\nu })=e^d\Lambda ^{\mu }_{\;\;\varrho }\otimes e^d\Lambda ^{\varrho }_{\;\;\nu } \label{w15} \\
&& \triangle (a^{\mu })=e^d\Lambda ^{\mu }_{\;\;\varrho }\otimes a^{\varrho }+a^{\mu }\otimes 1 \nonumber \\
&& \triangle (d)=d\otimes 1+1\otimes d \nonumber
\end{eqnarray}

Similar results hold for any inhomogeneous linear group (eg. the affine group). One can also try to apply this algorithm to 
the conformal group (see \cite{b6} for Lie algebra level version); however, the resulting algebra is so complicated 
that we were not able to put it in reasonable form.

As it was discussed in Ref. \cite{b9}, eq.(\ref{w2}) provides the proper framework for discussing the space-time symmetries 
of NCFT. In fact, any theory obtained from the Poincare (Weyl, conformal etc.) invariant commutative one by replacing the 
pointwise products by the star ones will be invariant under the action of quantum group counterpart of Poincare (Weyl, 
conformal etc.) group. Moreover, as indicated in Ref. \cite{b9}, the quantum group symmetry picture is equivalent to the more 
traditional one based on explicit symmetry breaking down to the stability subgroup of $\Theta ^{\mu \nu }$.\\

{\bf Appendix}\\

Let us remind simple facts concerning the geometry of invariant fields on Lie group manifold \cite{b14}. Let $G$\ be 
a Lie group with elements $g(\underline{a})$\ parametrized (locally at least) by $a^{\alpha }, \;\;\alpha =1,...,r$. Let 
further $\varphi (a,b)=(\varphi ^1(a,b),...,\varphi ^r(a,b))$\ be the composition function (we assume $\varphi ^{\alpha }
(a,0)=\varphi ^{\alpha }(0,a)=a^{\alpha },\;\;\; \alpha =1,2,...,r)$\
\begin{eqnarray}
g(\underline{a})g(\underline{b})=g(\varphi (\underline{a},\underline{b})) \label{w16}
\end{eqnarray}
Then the left - and right - invariant fields are defined in local coordinates by 
\begin{eqnarray}
X^L_{\alpha }=\mu ^{\beta }_{\;\;\alpha }{(\underline{a})}\frac{\partial }{\partial a^{\beta }}, \;\;\; X^R_{\alpha }=\tilde \mu ^
{\beta }_{\;\;\alpha }(\underline{a})\frac{\partial }{\partial a^{\beta }} \label{w17}
\end{eqnarray}
where
\begin{eqnarray}
&& \mu ^{\beta }_{\;\;\alpha }(\underline{a})=\frac{\partial \varphi ^{\beta }(\underline{a},\underline{b})}{\partial b^
{\alpha }}\mid _{b=0} \nonumber \\
&& \tilde \mu ^{\beta }_{\;\;\alpha }(\underline{a})=\frac{\partial \varphi ^{\beta }(\underline{b},\underline{a})}
{\partial b^{\alpha }}\mid   _{b=0} \label{w18}
\end{eqnarray}

The associativity law
\begin{eqnarray}
\varphi (\varphi (\underline{a},\underline{b}),\underline{c})=\varphi (\underline{a},\varphi (\underline{b},\underline{c}))
\label{w19}
\end{eqnarray}
differentiated with respect to $b^{\alpha }$\ at $\underline{b}=0$\ gives 
\begin{eqnarray}
\frac{\partial \varphi ^{\alpha }(\underline{a},\underline{c})}{\partial a^{\beta }}\mu ^{\beta }_{\;\;\gamma }
{(\underline{a})}=
\frac{\partial \varphi ^{\alpha }(\underline{a},c)}{\partial c^{\beta }}\tilde \mu ^{\beta }_{\;\;\gamma }{(\underline{c})}
\label{w20}
\end{eqnarray}

\end{document}